 \documentclass[acmsmall,nonacm]{acmart}
 \setcopyright{none}

\usepackage{enumerate}

\definecolor{CommentColor1}{rgb}{0.9,0.2,0.0}
\definecolor{CommentColor2}{rgb}{0.0,0.7,0.0}
\definecolor{CommentColor3}{rgb}{0.9,0.2,0.9}
\definecolor{CommentColor4}{rgb}{0.9,0.0,0.0}

\usepackage{mathtools} \usepackage{xspace} \newcommand{\FLIP}{\protect\reflectbox{F}LIP\xspace}
\AtBeginDocument{%
  \providecommand\BibTeX{{%
    \normalfont B\kern-0.5em{\scshape i\kern-0.25em b}\kern-0.8em\TeX}}}

\begin{document}

\title{Deep and Fast Approximate Order Independent Transparency}

\author{Grigoris Tsopouridis}
\authornote{All authors contributed equally to this research.}
\email{g.tsopouridis@uoi.gr}
\affiliation{%
  \institution{University of Ioannina}
  \country{Greece}
}
\author{Andreas A. Vasilakis}
\authornotemark[1]
\email{abasilak@aueb.gr}
\affiliation{%
  \institution{Athens University of Economics and Business}
  \city{Athens}
  \country{Greece}
}

\author{Ioannis Fudos}
\authornotemark[1]
\email{fudos@uoi.gr}
\affiliation{%
  \institution{University of Ioannina}
  \city{Ioannina}
  \country{Greece}
  \postcode{45110}
}

\renewcommand{\shortauthors}{Tsopouridis, et al.}

\begin{abstract}
  We present a machine learning approach for efficiently computing order independent transparency (OIT). Our method is fast, requires a small constant  amount of memory (depends only on the screen resolution and not on the number of triangles or transparent layers), is more accurate as compared to previous approximate methods, works for every scene without setup and is portable to all platforms running even with commodity GPUs. Our method requires a rendering pass to extract all features that are subsequently used to predict the overall OIT pixel color with a pre-trained neural network. We provide a comparative experimental evaluation and shader source code of all methods for reproduction of the experiments.
\end{abstract}



\begin{teaserfigure}
  \includegraphics[width=\textwidth]{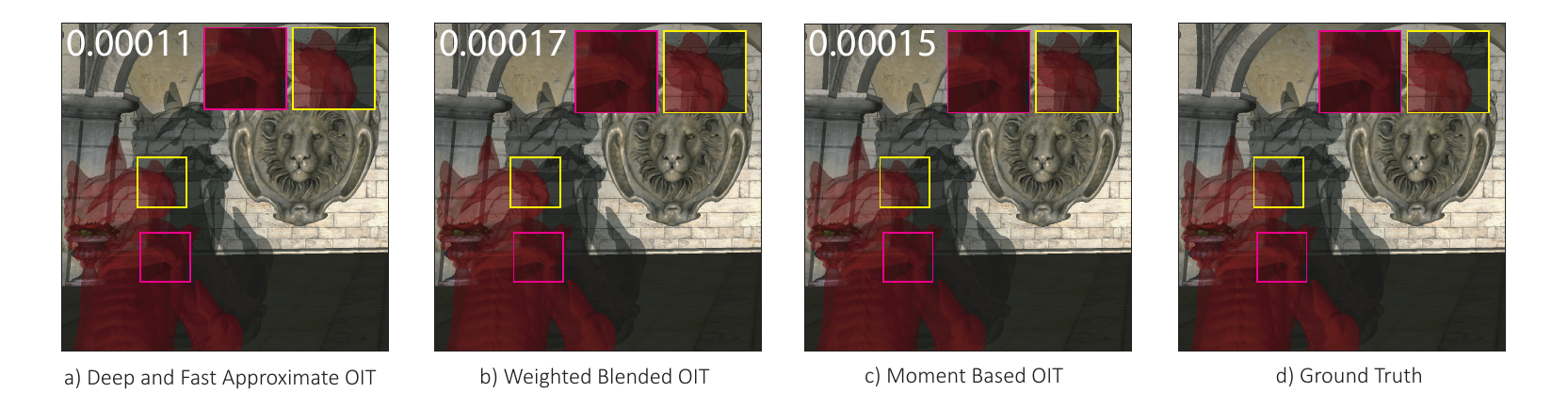}
  \caption{Our deep learning approach (a) produces visually convincing transparency results with respect to ground-truth images produced by an A-buffer \cite{yang2010} (d), qualitatively outperforming previous approximate methods \cite{McGuire2013,Munstermann2018,Sharpe2018} (b,c) using MSE as a metric.}
  \label{fig:teaser}
\end{teaserfigure}

\maketitle

\section{Introduction}
\label{sec:intro}

Order Independent Transparency (OIT) is a widely used technique in computer graphics that enables the rendering of transparent objects in a scene while maintaining the correct visual information~\cite{maule2011,Wyman2016}. Transparent objects in a scene, such as glass, water, or smoke, can often overlap, resulting in incorrect or unexpected rendering outputs. OIT techniques aim to solve this issue by ensuring that the correct ordering of transparent fragment samples is maintained~\cite{Vasilakis2020}, even when the objects intersect with each other. Specifically, OIT is derived by 
\begin{enumerate}
\item sorting all $n$ fragments of a pixel $p$ by their depth $z$:
$ [f_{n}, f_{n-1}, \dots, f_{1}]$,~so that $z_{n} \leq z_{n-1} \leq \dots \leq z_1$.
\item computing the pixel color $C_p=C_f(n)$ by the following recursive formula (also known  as \emph{over compositing operator}~\cite{PorterDuff1984}):
\begin{equation}
\label{eq:OVER}
C_f(i)=\begin{cases}
C_b & i = 0\\
a_iC_i + (1-a_i)C_f(i-1)&\ i = 1, \dots ,n
\end{cases}
\end{equation}
where $C_b$ is the background color, $C_i$ and $a_i$ are the color and opacity of fragment $f_i$ and $C_f(i)$ is the color in front of fragment $f_i$. Therefore, $C_f(n)$ is the color of pixel $p$.
\end{enumerate}

The first step requires to store all fragment information in a per pixel buffer and then to sort all fragments according to their depth value. To achieve real-time performance several algorithms have been used that maintain a memory space per pixel, where information regarding \emph{all} fragments for this pixel is stored. This rendering pipeline is known as A-buffer~\cite{Carpenter1984} and can be implemented either by allocating a predetermined memory portion per pixel that can accommodate all fragments~\cite{Liu2010} which is fast but memory demanding, or by creating per pixel linked-lists~\cite{yang2010,Vasilakis2012}, which is slower but allocates exactly the required memory space. 

To approximate OIT with interactive performance and reasonable memory requirements the following categories of approaches have been explored:
\begin{itemize}
    \item[(a)] {\em average OIT}  approaches \cite{Meshkin2007,Bavoil2008OrderIT} utilize only global average information such as average opacity and average color. They work on the pixel level, are fast but lack accuracy as compared to $k$-buffer and blended approaches.
    \item[(b)] {\em blended OIT} approaches \cite{McGuire2013,Munstermann2018,Sharpe2018,Friederichs2020} use information which is passed on by each fragment and then is averaged or summed at the pixel level. The processing by each fragment is conducted via a blending function. They are fast and provide good accuracy in the expense of a small processing overhead per fragment.  
    \item[(c)] {\em $k$-buffer and multifragment} OIT approaches \cite{Salvi2011, Wyman2016} use a $k$-subset selection of all generated fragments. The $k$ value can be fixed, variable or adaptive and may refer to $k$ closest or a random subset of fragments.
    They sacrifice performance and memory to offer the best quality approximation working with relatively advanced GPUs.
    \item[(d)] {\em hybrid OIT} approaches \cite{Maule2013,Salvi2014,Tsopouridis2022} combine  $k$-buffer approaches with average or blended OIT approaches for the rest of the fragments that are not part of the $k$-buffer set (also known as the \emph{tail} part). 
\end{itemize}

In this work, we present a deep learning approach to robustly compute OIT that outperforms all previous blending and average OIT approaches. Our Deep and Fast Approximate OIT (DFAOIT) receives as input features, global per pixel measures such as average color and opacity. Using this input, a pre-trained neural network outputs the final pixel transparency color at once. The neural network inference has been implemented to run efficiently on a fragment shader, without requirements for state of the art graphics hardware. 


The approach introduced in this work makes the following technical contributions: 
\begin{itemize}
\item To our knowledge, this is the first approach to use deep learning for approximating OIT with global per pixel measures. 
\item Our approach offers comparable performance with approximate average or blended OIT approaches and offers $20\%-80\%$ better accuracy (using the mean square pixel color error) as compared to the most competent previous approaches. 
\item We offer a cross-platform implementation for OIT rendering on any device, where no hardware that guarantees a deterministic pipeline order is required.
\item Our method requires no parameters to be determined or tuned by the user before rendering.
\item DFAOIT can be used for improving the tail part of hybrid OIT approaches.
\end{itemize}

The structure of the rest of this paper is as follows. Section \ref{sec:related} provides comparison with state of the art. Section \ref{sec:method} outlines our DFAOIT method and Section \ref{sec:implementation} offers implementation details. Section \ref{sec:experiments} conducts a thorough comparative experimental evaluation. Finally, Section \ref{sec:conclusions} offers conclusions and future research directions. 

\section{Related Work}
\label{sec:related}

\noindent
\textbf{Rendering Transparency}.
Many games and simulations use transparent objects like glass, leaves, hair or fog to add realism and depth to their environments~\cite{Moore2009,Martin2018,Sikachev2021}. Existing techniques often require separating opaque and transparent geometry into separate rendering passes and the respective output buffers are merged together to form the final blended color~\cite{Schollmeyer2015}. Traditionally, a standard set of compositing operators is applied to combine transparent surfaces~\cite{PorterDuff1984}. However, this process is challenging since it requires an order-dependent processing on a per-geometry basis (per frame) to achieve the desired blending result~\cite{Han2018}. Moreover, an inaccurate solution in cases of intersecting or circularly overlapping geometry is also possible. Advances in graphics hardware have fuelled an explosion of research and development in the field of order-independent transparency~\cite{maule2011,Wyman2016} to approximate alpha compositing in screen space without explicit geometry ordering. Without loss of generality, these methods can be classified into two board categories: (i) \emph{blended} (categories (b) and (c) of Sec. \ref{sec:intro}) and (ii) \emph{multifragment} transparency rendering methods (categories (a) and (d) of Sec. \ref{sec:intro}). Despite the recent trends of neural rendering~\cite{Tewari2020}, and its application to enhance screen-space global illumination~\cite{Nalbach2017,Thomas2017,Zhang2020,Datta2022}, to our knowledge we are the first to exploit deep learning for efficiently computing OIT effects.\\

\noindent
\textbf{Blended transparency rendering}. 
Real-time 3D applications, such as games, require acceptable transparency results within a limited computation time cost while respecting the current rendering  graphics engine pipeline. Blended OIT methods focus on portability and speed by adjusting the over compositing operator (Eq.~\ref{eq:OVER})~\cite{PorterDuff1984} so that its arguments commute.\emph{Weighted-sum alpha blending}~\cite{Meshkin2007} was the first to introduce blended OIT by redefining the blending formulation as follows:
\begin{equation}
\label{eq:WSUM}
C_p = \textstyle\sum_{i-1}^{n} C_ia_i + C_0(1-\textstyle\sum_{i=1}^{n} a_i)
\end{equation}
While the alpha blending equation was simplified in order to ignore the order-dependent terms, this method produces unacceptable quality (overly dark or bright color) for high opacity values. Following the same idea,
\emph{Weighted-average alpha blending}~\cite{Bavoil2008OrderIT} offers a better OIT approximation by distributing the transmittance equally among the transparent fragments:
\begin{align}
\label{eq:WAVG}
C_p     &= C_{avg}(1-(1-a_{avg})^{n}) + C_0(1-a_{avg})^{n}\\
C_{avg} &= \frac{\textstyle\sum_{i=1}^{n} C_ia_i}{\textstyle\sum_{i=1}^{n}a_i},~\text{and }
a_{avg}  = \frac{\textstyle\sum_{i=1}^{n}a_i}{n}
\end{align}
The blending average method overcomes color saturation exhibited in previous method, but comes with poor transparency output on scenes with high opacity values (e.g. avoids to properly support invisible surfaces).
\emph{Weighted-blended alpha blending}~\cite{McGuire2013} extended the previous operator to compute exact coverage of the background, which also improves the color term. To create an occlusion cue between transparent surfaces, the operator was further color-weighted to include depth information.
The weights for each fragment $f_i$ are computed from a monotonic decreasing function $w(z_i,a_i)$ (several proposed in the original paper). The operator can be formulated as: 
\begin{align}
\label{eq:WBOIT}
C_p      &= C_{wavg}a_{net} + C_0(1-a_{net})\\
C_{wavg} &= \frac{\textstyle\sum_{i=1}^{n} C_ia_iw(z_i,a_i)}{\textstyle\sum_{i=1}^{n}a_iw(z_i,a_i)},~\text{and }
a_{net}   = 1- \textstyle\prod_{i=1}^{n}(1-a_i)
\end{align}

In all cases, weights have to be carefully tuned for each scene in order to provide plausible results. Recently, Friederichs et al. extended the previous algorithm via a layering approach~\cite{Friederichs2020}. They divided the view frustum into multiple depth intervals on which the weighted-blended OIT algorithm is applied separately leading to more robust depth cues and improves color precision. However, it requires excessive memory demands due to linked lists construction~\cite{yang2010} and it struggles to deliver good performance when a high number of bins is required. 

Moment transparency~\cite{Sharpe2018,Munstermann2018} achieves better results in real-time by replacing the empirically determined weights $w(z_i,a_i)$ with \textit{moments} to capture and reconstruct the transmittance function. Several variants (number of power moments, overestimation, low-precision buffers) were also introduced on each with trade-offs between quality, performance, and robustness.

 \noindent
\textbf{Multifragment transparency rendering}.
There has been a significant increase in the number of techniques that use multiple fragments to create realistic transparency~\cite{Vasilakis2020}. These methods have the advantage of compositing more geometry layers, by preserving more information from the fragment sampling domain. The multifragment process can be classified into two categories based on the maximum number of fragment samples per pixel captured in a single iteration: (i) \emph{depth-peeling} and (ii) \emph{A-buffer}.

Depth-peeling methods iteratively extract fragment information in a depth-sorted manner. Depending on the specific algorithm, each rendering iteration extracts a batch of fragments with a guaranteed depth order. This process continues until a sufficient visual quality has been achieved or all fragments have been extracted. 
The most prominent method is probably the \emph{front-to-back depth-peeling}~\cite{Everitt2001}, which involves rendering the geometry multiple times and extracting a \emph{single} fragment layer in ascending depth order for each iteration. While several extensions have been introduced to speed up the fragment peeling process by extracting multiple fragments per iteration~\cite{Bavoil2008OrderIT,Liu2009,Vasilakis2013,Mara2016},
depth-peeling methods fail to behave interactively on complex environments since the geometry is repeatedly rasterised multiple times. 
By exploiting more fragment storage for less rendering passes,
\emph{multilayer depth peeling} (also well-known as \emph{k-buffer}~\cite{Bavoil2007} methods) can be considered as the most preferred framework for optimal fragment \textit{k}-subset selection. These approaches accumulate transparent geometry with the help of a fixed number of layers~\cite{Salvi2011,Salvi2014,Wyman2016}. To approximate
the ground-truth OIT solution, a fragment merging strategy is performed in a streaming fashion. The original methods required specialized hardware modifications in pixel synchronisation~\cite{Salvi2013, Vasilakis2014} that were made available later on, limiting the portability of these methods. Finally, Maule et al.~\cite{Maule2013} introduced a \emph{hybrid transparency} method, in which the closest transparent layers are accurately composited using a \textit{k}-buffer, while the remaining ones are quickly approximated using blended transparency techniques. This allows for a balance of accuracy and efficiency in the compositing process. Recently, Tsopouridis et al. improved this work by dynamically distributing more fragments in pixels that contribute more to the visual result guided by a deep learning prediction mechanism~\cite{Tsopouridis2022}.

Historically, \emph{A-buffer} method has been the best solution for handling multiple transparent layers of rasterized geometry~\cite{Carpenter1984}. It uses multiple per-pixel depth-order fragment linked lists to accurately simulate transparency on the GPU~\cite{yang2010,Vasilakis2012}. However, one of the main drawbacks of A-buffer-based solutions is their high memory requirements. Because the number of fragments generated during the rendering process is unpredictable, the memory needed to store them can quickly become overwhelming, making A-buffer unsuitable for interactive applications on constrained rendering devices~\cite{Vasilakis2020}.
\section{Deep and Fast Approximate OIT}
\label{sec:method}
To derive a formula for approximating the OIT color $C_f(n)$ of the entire pixel $p$, by average fragment quantities, we expand Eq. \ref{eq:OVER}:  
\begin{equation}
\label{eq:oit-expanded}
\begin{split}
C_f(n) = &a_{n} C_{n} + (1-a_{n}) a_{n-1}C_{n-1} + \dots  \\ 
& + a_1 C_{1} \prod_{i=2}^{n}{(1-a_i)}  +  C_{b} \prod_{i=1}^{n}{(1-a_i)} 
\end{split}
\end{equation}
and then we derive the approximation:
\begin{equation}
 \label{eq:oit-approximate2}
 \begin{split}
\widehat{C}_f(n) = \, & g(n,1) 
+ h(a_{avg}, C_{avg}, C_{acc}, g(n,1)) \, \\  
& + C_b \prod_{i=1}^{n}(1-a_i) 
\end{split}
\end{equation}
with
$
g(i,k) = a_i C_i + (1-a_i)a_{i-k}C_{i-k}
$, 
where $C_{avg}$ and $a_{avg}$ is the average color and average opacity of the farthest $n-2$ fragments of pixel $p$, $C_{acc}$ is the sum of the premultiplied alpha colors $\sum_{i=1}^{n} a_i C_i$.  
Finally, $g(n,1)$ is the combined color of the first two fragments and $h$ is a function that approximates the sum of terms $$\sum_{i=1}^{n-2}a_iC_i\prod_{k=i+1}^{n}{(1-a_k)}$$
For example, for $n=4$ where we have four fragments, we may use the following function $h(x,y,z,w)= (1-x)^2 (z-w-2x^2y)$.
The following substantiates that this choice for function $h$ provides a valid approximation.
\begin{equation*}
 \begin{split}
\widehat{C}_f(4) = & \, a_4 C4 + (1-a_4)a_3 C_3 +\\ 
&(1-a_{avg})^2(a_2 C_2 + a_1 C_1 + a_4 a_3 C_3 - 2a_{avg}^2 C_{avg})\\ 
&+ (1-a_4)(1-a_3)(1-a_2)(1-a_1)C_b\\
\\ \approx & \, a_4 C4 + (1-a_4)a_3 C_3 +\\ 
&(1-a_{avg})^2(a_2 C_2 + a_1 C_1 - a_1 a_2 C_1 +a_4a_3C_3 -a_4a_3C_3)\\ 
&+ (1-a_4)(1-a_3)(1-a_2)(1-a_1)C_b\\
\\ \approx & \, a_4 C4 + (1-a_4)a_3 C_3 +\\ &(1-a_4)(1-a_3)(a_2 C_2 + (1-a_2)a_1 C_1) + \\
&(1-a_4)(1-a_3)(1-a_2)(1-a_1)C_b\\
= &~C_f(4)
\end{split}
\end{equation*}

However, since it is not straightforward how to derive a closed formula for function $h$ that will be valid for any value of $n$, we revert to a machine learning approach to determine an approximating function $h$.
Therefore, we select as initial candidate input features the number of fragments, the average opacity and color of the farthest $n-2$ fragments, the sum of $a_i C_i$, and the combined color of the first two fragments. The term of the background color can always be added at the final step using the exact formula: $C_b \prod_{i=1}^{n}(1-a_i)$.

In the rest of this section, we describe the training and inference phases of our deep and fast OIT technique. We introduce a “one network for one effect” approach for training the network, that maps the per-pixel inputs to an OIT RGB color.
We first describe the neural network architecture (Sec.~\ref{sec:nn}) along with the selection of its input features and training parameters. Then, for the training phase (Sec.~\ref{sec:training}), we explain how we acquire the necessary screen-space training data for the supervised learning process.
Finally, we discuss the limitations of various network architectures and input features that are not utilized by our method (Sec.~\ref{sec:ablation}).

\subsection{Network Architecture and Feature Selection}
\label{sec:nn}

\begin{figure}[!b]
	\centering
	\includegraphics[scale=1.00]{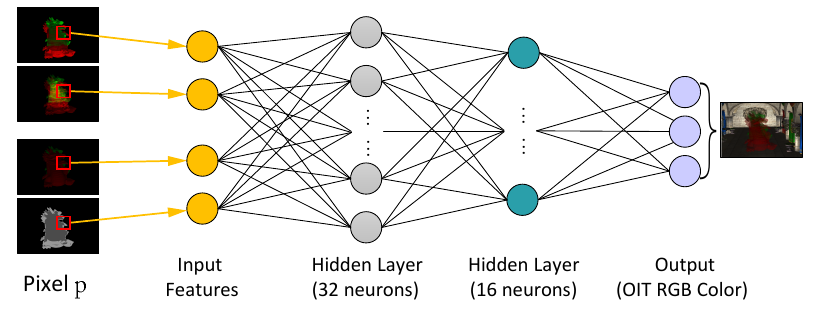}
	\caption{A visualization of our neural network that maps the input features of pixel $p$ into an OIT RGB color output.}
	\label{fig:nn_architecture}
\end{figure}
We have selected a compact fully connected multilayer perceptron to predict per-pixel OIT RGB colors, given ten float input features (Fig.~\ref{fig:nn_architecture}). 
The neural network is composed of an input, an output, and two hidden layers. 
The hidden layers are composed of 32 and 16 neurons respectively, and use ReLU as their activation function. The output layer returns three float values $C_p = (r_p,g_p,b_p) \in{}[0, 1]$ that represent the predicted OIT RGB color, using the sigmoid activation function. For training the network, Mean-Squared Error (MSE) is used as a loss function and Adam \cite{adam} with a learning rate of $0.001$ as an optimizer. Finally, we train our network for 3000 epochs using a batch size of 4096. All input features are in the range of [0,1], except the accumulated premultiplied color.

The neural network expects the following 10 per-pixel float features as input:
\begin{itemize}
    \item Correct OIT RGB color of the two front fragments: \\ $g(n,1)=a_nC_n + (1-a_n)a_{n-1}C_{n-1}$.
    \item Average opacity, excluding the front two fragments: \\ $a_{avg}=\textstyle\sum_{i=1}^{n-2} a_i/n$.
    \item Average RGB color, excluding the front two fragments: \\
    $C_{avg}=\textstyle\sum_{i=1}^{n-2} C_i/n$.
    \item Accumulated premultiplied alpha RGB color:\\ $C_{acc}=\textstyle\sum_{i=1}^{n} a_iC_i$.
\end{itemize}

In most cases, the furthest a fragment is from the camera, the smaller impact it has on the final color (as derived by Eq.~\ref{eq:OVER} and \ref{eq:oit-expanded}). Consequently, the input features must provide color information throughout the whole depth of a pixel to enable the neural network to map them to an OIT color. The exact OIT color of the first two fragments is vital to the final OIT color and it allows the neural network to better approximate the exact OIT and retain finer details. The average color $C_{avg}$ and opacity $\alpha_{avg}$ provide the contribution of the remaining fragments to the final color, playing a complementary role to $g(n,1)$. Finally, the accumulated alpha-multiplied color $C_{acc}$ gives a general insight about the color and opacity of the pixel. All these features will assist in determining function $h$ of Equation \ref{eq:oit-approximate2}. 

\begin{figure}[!t]
	\centering
	\includegraphics[scale=1]{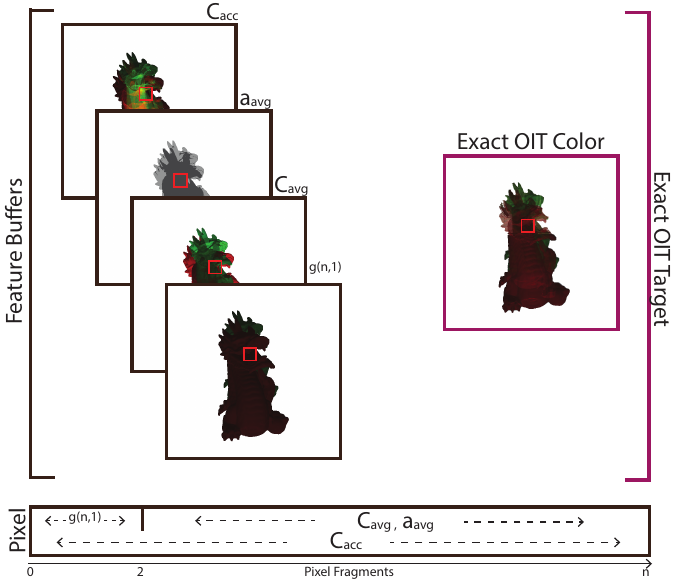}
	\caption{Visualizing training pairs (up). The per-pixel input features are generated using rasterization. The targets can be generated correctly by A-buffer~\cite{yang2010}. The input features aim to gather color information throughout the entire pixel (down).}
	\label{fig:features}
\end{figure}

\subsection{Training Dataset}
\label{sec:training}
We trained our neural network using supervised learning, mapping the per-pixel input features to an RGB color. The training examples are generated by exporting the information from all generated fragments (e.g. an A-buffer~\cite{yang2010}) and computing the input features as well as the exact OIT RGB color (target). 
To this end, we created a dataset that contains pairs of per-pixel inputs and the target per-pixel exact OIT RGB color (Fig.~\ref{fig:features}). This dataset includes input features from several fully transparent scenes of varying depth complexity (10 to 50 fragments). Each scene contains a $single$ multi-colored 3D object that has a random {\em per-pixel} global opacity $\alpha \in [0.01, 0.85]$, ensuring that our network can generalize over a large range of opacity values. We used 8 different scenes, totalling five million training examples. The training dataset plays a major role in determining the final OIT result, as examples with different colors, depth complexities and opacities must be included.

\subsection{Ablation Study}
\label{sec:ablation}
As our method targets real-time performance and relatively low memory requirements, we were limited in terms of network architectures and input features. In this section, we discuss the limitations of various other network architectures and features. Regarding network architectures, we discarded the employment of CNN early on, as while using a CNN may be beneficial in terms of quality due to its ability to correlate neighboring pixel information, it is quite demanding in terms of computations and memory even when modern hardware is exploited~\cite{Datta2022} and hard to implement in a fragment shader. Additionally, our experiments have shown that using larger hidden layers (up to 256 neurons) slightly improve overall accuracy but are not worth the significant performance and memory overhead. For example, doubling the layer size (64 and 32 neurons) only improved the quality by $6\%$, but doubled the weights memory cost, as they need to be stored in a shader, and inference time. 

Neural network inputs should be fast and easy to compute (even by commodity graphics hardware). In a way, our method splits pixel space into two parts, the exact front OIT part $C_{e}$, and the supplementary approximation tail part $C_{acc}, C_{avg}, a_{avg}$. To improve quality, we can better represent color changes in pixel space by further splitting the depth of the pixel into three or four (or more) depth buckets (inspired by Layered Weighted Blended Order-Independent Transparency~\cite{Friederichs2020}), with the first bucket containing exact OIT information, and the remaining buckets containing average color and opacity, each bucket should cover an increasingly larger depth partition, as farthest fragments do not contribute much to the final color. While this could potentially increase quality, it is a challenge to find optimal ways to split space, and it greatly increases memory and performance overhead, deeming it unsuitable for our method. Additionally, it was possible to include the exact color of more than two fragments. However, we consider this unsuitable for our method as it increases memory and computational requirements without significant accuracy gain. Finally, including pixel depth information (mean, min, max) as an input feature did not increase the quality of our predictions, however, introducing both local (pixel) and global (scene) features could potentially resolve this problem with a performance overhead, providing a cue regarding fragment ordering in the scene. In the future, one could possibly employ machine learning for deriving a \emph{color weight} function $f(z_i,C_i)$, better representing the occlusion of objects ~\cite{McGuire2013}.
\section{Implementation}
\label{sec:implementation}

\begin{figure}[!t]
	\centering
	\includegraphics[scale=0.65]{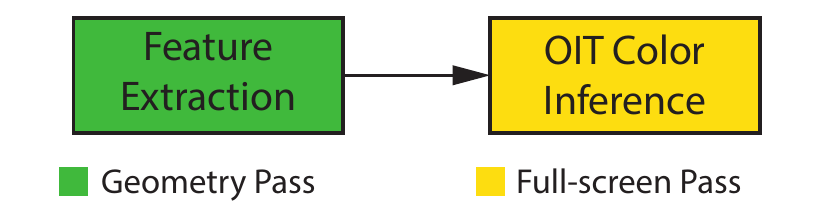}
	\caption{Rendering flow diagram of our method.}
	\label{fig:method}
\end{figure}

Our method requires two passes (Fig. \ref{fig:method}), (i) a geometry pass for \textit{Feature Extraction}, that computes and stores the neural network input features, and (ii) a full-screen pass for \textit{OIT Color Inference}, that uses the input features and the stored weights and biases of the neural network to compute the predicted RGB OIT color. \textit{Feature Extraction} accumulates the color, premultiplied color, opacity as well as the color and opacity of the first two fragments. We extract the two closest to the camera fragments using a custom depth comparison implementation with interlocks, however other, more advanced ways can be employed such as $k$-buffer \cite{Bavoil2007, Vasilakis2014}, depth-peeling \cite{Everitt2001} or by utilizing minimum separation and temporal information \cite{Mara2016}. 

\begin{figure}[!t]
	\centering
	\includegraphics[]{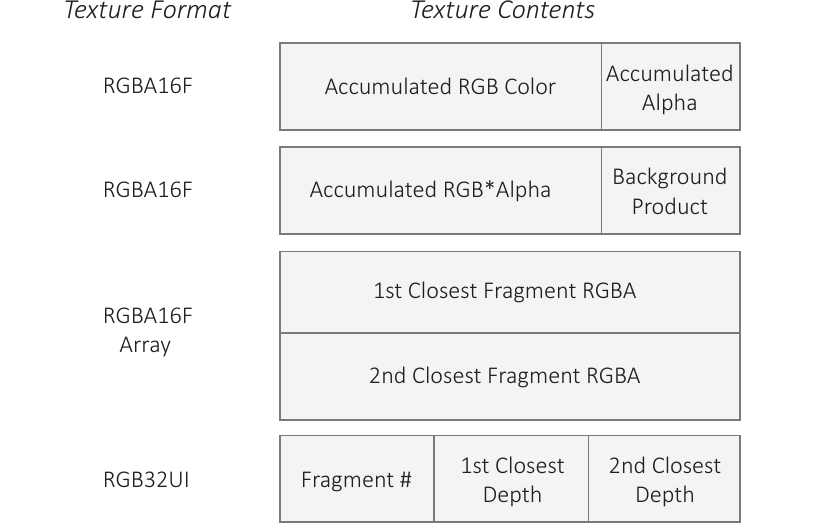}
	\caption{Texture contents and sizes of our method.}
	\label{fig:textures}
\end{figure}

The neural network approximates the color $C_t$ of transparent objects, excluding the contribution of the background color $C_b$, which is accounted for in \textit{OIT Color Inference} using the background product. Thus, the final color $C_p$ is computed as follows:
\begin{equation}
\label{eqn:weight}
C_p = C_t + C_b\prod_{i=1}^{n}(1-a_i) 
\end{equation}
Our method requires four textures (Fig. \ref{fig:textures}):
\begin{itemize}
    \item an RGB 32-bit unsigned integer texture that contains the number of fragments per pixel, which is used to compute the average colors and opacity, and the depths of the two closest fragments.
    \item an RGBA texture, using 16-bits per value, that accumulates color and alpha values.
    \item an RGBA texture, using 16-bits per value, that accumulates colors multiplied by their opacities ($a_iC_i$) and the background product $\prod_{i=1}^{n}(1-a_i)$.
    \item a two-layered RGBA texture that uses 16-bits per value, storing the colors of the two closest fragments.
\end{itemize} 

To provide a fair comparative experimental evaluation (Sec.~\ref{sec:experiments}), we implemented Weighted Sum (WSUM) \cite{Meshkin2007} using an RGBA texture with 32 bits per value, that accumulates color and alpha values. Both Weighted Blended (WBOIT) \cite{McGuire2013} and Weighted Average (WAVG) \cite{Bavoil2008OrderIT}  allocate two textures, an RGBA accumulation texture with (32-bits per value) that accumulates color, and an 32-bit unsigned int texture that stores the background product ($\prod_{i=1}^{n}(1-a_i)$, WBOIT) and the number of fragments per pixel (WAVG). Hybrid Transparency (HT) \cite{Maule2013} requires a two layered 16-bit RGBA texture to store the two closest fragment colors and opacities, a 32 bit RGBA texture to accumulate colors and opacities which is used for the tail color computation and, a 32-bit RG texture that stores the background coefficient and the number of per pixel fragments. Moment Based Order-Independent Transparency (MBOIT) \cite{Sharpe2018,Munstermann2018}, requires a 32-bit RGBA texture to store blended color, a 32 bit RGBA texture to accumulate moments and, a 32-bit RG texture to accumulate the total absorbance and the normalization value. 
Our method sacrifices more memory budget (352 bits per pixel), compared to the competition (WBOIT, WAVG: 150 bits per pixel - WSUM: 64 bits per pixel - HT, MBOIT: 320 bits per pixel), for providing superior visual effects  with real-time performance (Sec.~\ref{sec:experiments}). 

All methods were implemented in OpenGL 4.6, using interlocks and atomic operations where needed and similar texture sizes of 32-bit float or unsigned integer per value. Note that all methods, including ours, can be easily ported to older OpenGL versions, OpenGL ES and WebGL via core GPU-accelerated \emph{frame buffer blending operations}, thus supporting older GPUs and special purpose software platforms. In the low-end hardware variant of our method, depth-peeling is first used to extract the two closest fragments to the camera and an additional geometry pass is then performed to count the per-pixel fragments and accumulate the color and opacity neural network input features.
Nevertheless, we observed that for our method there is no performance gain or loss when the low-end hardware variant is used. Shader source code of this work, for both low- and high-end graphics hardware, is available online \footnote{https://github.com/gtsopus/DFA}.

\section{Experimental Evaluation}
\label{sec:experiments}

\begin{figure*}[!ht]
	\centering
	\includegraphics[width=\textwidth]{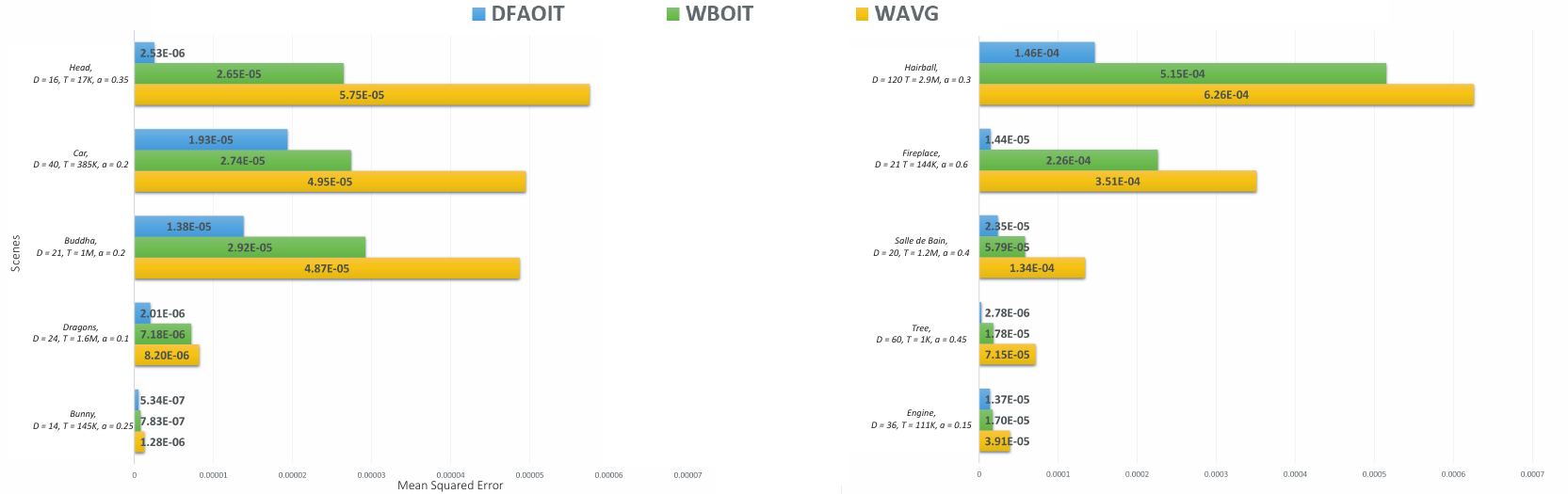}
	\caption{MSE measurements of the three OIT approaches on several scenes with varying depth complexity $D$, triangle number $T$, and opacity $a$. Our method outperforms all methods in all cases, offering better image quality.}
	\label{fig:graph}
\end{figure*}

\begin{figure*}[h]
	\centering
	\includegraphics[width=\textwidth]{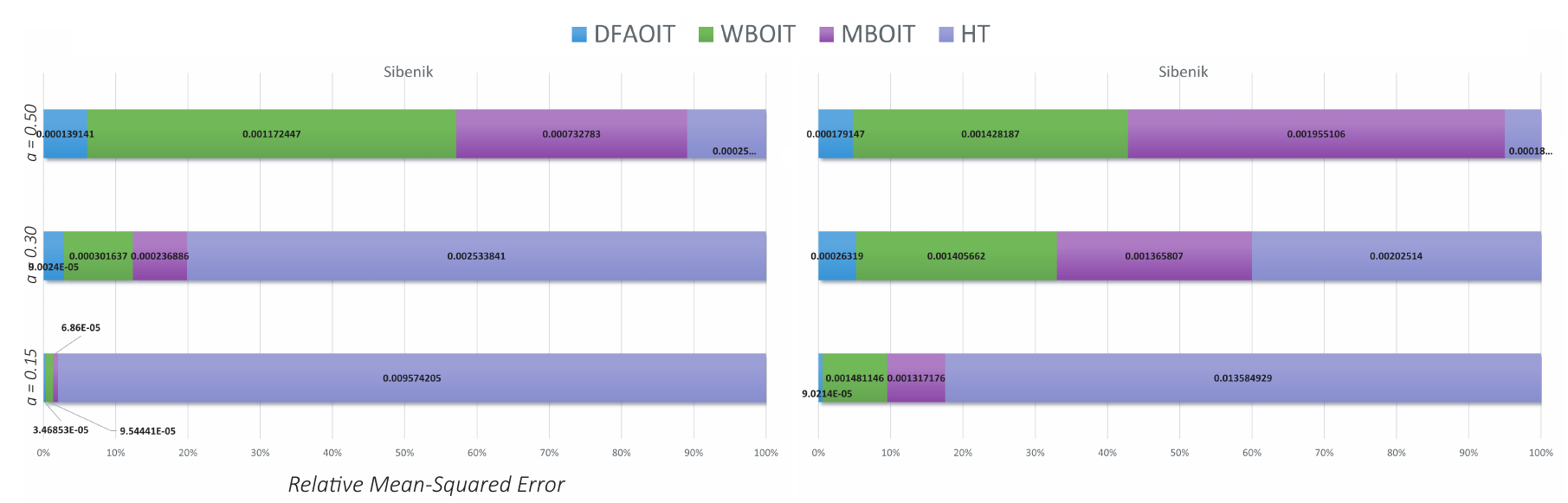}
	\caption{Relative MSE measurements of four OIT approaches on two scenes (Sponza: $T=262K$, $D=40$, Sibenik: $T=76K$, $D=21$) with varying opacity values $a$. We observe that our method yields a better result independent from the opacity values.}
	\label{fig:graph_rel}
\end{figure*}

\begin{figure*}[h]
	\centering
	\includegraphics[width=\textwidth]{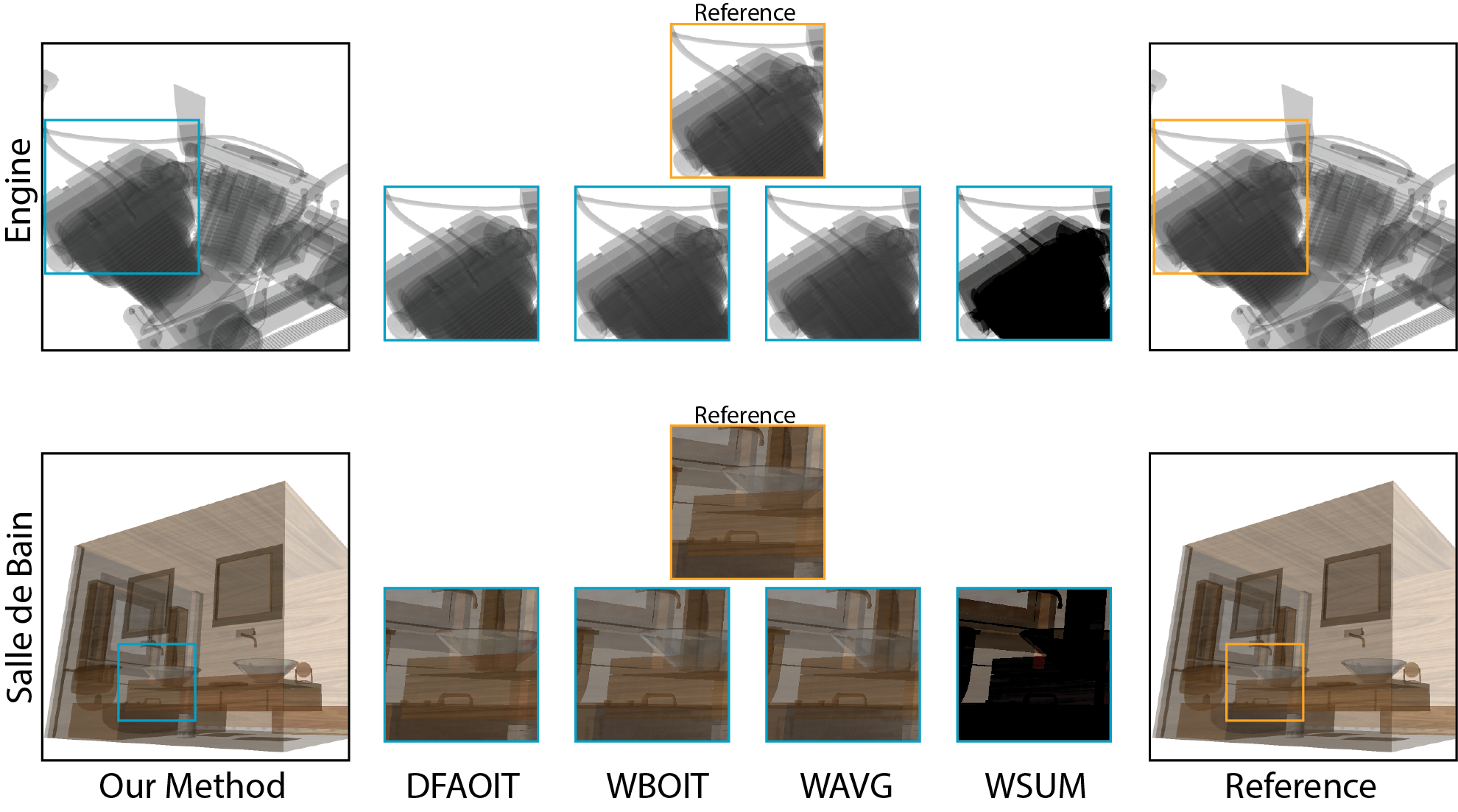}
	\caption{A quality evaluation between the four OIT approaches. Our method achieves better quality results by retaining higher transmittance detail. (Engine: $T=144K$, $D=21$, Salle de Bain: $T=1.2M$, $D=20$)}
	\label{fig:graph_qcomp1}
\end{figure*}

We present an experimental evaluation of our method (DFAOIT) as compared to WBOIT~\cite{McGuire2013}, WAVG~\cite{Bavoil2008OrderIT}, WSUM~\cite{Meshkin2007}, HT~\cite{Maule2013} and, MBOIT~\cite{Sharpe2018} with regards to performance and image quality. We have performed experiments on several scenes, with varying colors, depth complexity and opacity using a $1920\times1080$ viewport for all experiments except those of Table \ref{tab:resolutions}. All experiments were conducted on a system with NVIDIA RTX 2080 Super GPU. 

We have conducted experiments using several scenes that differ with regard to color and depth complexity, resulting in diverse input features. The test scenes were not included in the training dataset. As our method follows the "one network for one effect approach" it is expected to work in every environment without need of any further fine-tuning, similarly to WAVG, HT and WSUM. However, WBOITs \cite{McGuire2013} weights may require fine-tuning, a tedious process that has a significant impact on results. For this reason, for WBOIT we chose one of the author recommended depth weights: $w(z,a) = a\cdot \max(10^{-2},3\cdot 10^3(1-z)^3)$, that fits the requirements, and tuned to derive better results for our scene and camera setup. In general, our method outperforms the competing methods in terms of quality while maintaining real-time performance. The ground truth images were generated using a per-pixel linked-lists A-buffer implementation~\cite{yang2010}. Our method was also implemented using blending operations, offering similar results. \\

\noindent
\textbf{Performance}. As illustrated in Table \ref{tab:performance}, our method performs slightly slower than WBOIT, offering a middle ground between WBOIT and MBOIT, achieving real-time performance in scenes of varying depth complexity, with only a minor performance overhead that depends on the collection of the network input features and the inference time of the neural network (Fig.~\ref{fig:method}). Our method scales and performs well in multiple resolutions (Table \ref{tab:resolutions}), with the feature extraction stage depending also on the number of scene fragments and, the neural network inference time depending on the number of transparent pixels. \\
 
\begin{table}[!t]
\caption{DFAOIT exhibits a slightly increased, but comparable, performance behavior compared to prior art, in scenes with varying depth complexity D. WSUM and WAVG perform better but are omitted since they are not competent with regard to accuracy according to the quality experiments.}
\label{tab:performance}
\begin{tabular}{|lllll|}
\hline
\multicolumn{1}{|l|}{\textbf{Method}} & \multicolumn{1}{l|}{HT}   & \multicolumn{1}{l|}{WBOIT} & \multicolumn{1}{l|}{DFAOIT} & MBOIT \\ \hline
\multicolumn{5}{|l|}{1. \textit{Two Dragons, D = 24}}                                                                                         \\ \hline
\multicolumn{1}{|l|}{Time (ms)}       & \multicolumn{1}{l|}{1.3}  & \multicolumn{1}{l|}{1.6}   & \multicolumn{1}{l|}{1.7}    & 2.3   \\ \hline
\multicolumn{5}{|l|}{2. \textit{Hairball, D = 120}}                                                                                           \\ \hline
\multicolumn{1}{|l|}{Time (ms)}       & \multicolumn{1}{l|}{27.3} & \multicolumn{1}{l|}{29.4}  & \multicolumn{1}{l|}{32.1}   & 45.6  \\ \hline
\multicolumn{5}{|l|}{3. \textit{Car, D = 40}}                                                                                                 \\ \hline
\multicolumn{1}{|l|}{Time (ms)}       & \multicolumn{1}{l|}{1.9}  & \multicolumn{1}{l|}{2.1}   & \multicolumn{1}{l|}{2.5}    & 3.7   \\ \hline
\multicolumn{5}{|l|}{4. \textit{Engine, D = 36}}                                                                                              \\ \hline
\multicolumn{1}{|l|}{Time (ms)}       & \multicolumn{1}{l|}{2.5}  & \multicolumn{1}{l|}{2.6}   & \multicolumn{1}{l|}{2.9}    & 3.8   \\ \hline
\multicolumn{5}{|l|}{5. \textit{Salle de Bain, D = 30}}                                                                                       \\ \hline
\multicolumn{1}{|l|}{Time (ms)}       & \multicolumn{1}{l|}{1.6}  & \multicolumn{1}{l|}{1.8}   & \multicolumn{1}{l|}{2.2}    & 2.8   \\ \hline
\multicolumn{5}{|l|}{6. \textit{Sibenik}, D = 21}                                                                                             \\ \hline
\multicolumn{1}{|l|}{Time (ms)}       & \multicolumn{1}{l|}{2.0}  & \multicolumn{1}{l|}{2.7}   & \multicolumn{1}{l|}{3.0}    & 3.9   \\ \hline
\end{tabular}
\end{table}

\begin{table}[thb]
\caption{Benchmarks of each stage of our method in different resolutions. DFOIT performs well in scenes with varying depth complexity D. F.E.T. is the feature extraction time and I.T. is the inference time.}
\label{tab:resolutions}
\begin{tabular}{|lllll|}
\hline
\multicolumn{1}{|l|}{\textbf{Resolution}}              & \multicolumn{1}{l|}{\small\textit{1280x720}} & \multicolumn{1}{l|}{\small\textit{1920x1080}} & \multicolumn{1}{l|}{\small\textit{2880x1620}} & \small\textit{3840x2160} \\ \hline
\multicolumn{5}{|l|}{\textit{Sponza, D = 40}}                                                                                                                                                            \\ \hline
\multicolumn{1}{|l|}{\textit{F.E.T. (ms)}} & \multicolumn{1}{l|}{1.6}               & \multicolumn{1}{l|}{3.0}                & \multicolumn{1}{l|}{4.4}                & 8.5                \\ \hline
\multicolumn{1}{|l|}{\textit{I.T. (ms)}}     & \multicolumn{1}{l|}{1.7}               & \multicolumn{1}{l|}{2.3}                & \multicolumn{1}{l|}{3.1}                & 3.7                \\ \hline
\multicolumn{5}{|l|}{\textit{Sibenik, D = 21}}                                                                                                                                                           \\ \hline
\multicolumn{1}{|l|}{\textit{F.E.T (ms)}} & \multicolumn{1}{l|}{1.1}               & \multicolumn{1}{l|}{1.7}                & \multicolumn{1}{l|}{2.3}                & 3.2                \\ \hline
\multicolumn{1}{|l|}{\textit{I.T. (ms)}}     & \multicolumn{1}{l|}{1.0}               & \multicolumn{1}{l|}{1.4}                & \multicolumn{1}{l|}{2.0}                & 2.8                \\ \hline
\end{tabular}
\end{table}

\noindent
\textbf{Quality}. Our method provides the best image quality in terms of mean squared error (MSE) compared to the aforementioned competing average (WAVG) and earlier blending OIT approaches (WBOIT) (Fig. \ref{fig:graph}) and compared to modern blending (MBOIT) and hybrid (HT) approaches (Fig. \ref{fig:graph_rel}). \textit{WSUM} produces the lowest quality images as the alpha blending equation is over-simplified and ignores the order-dependent terms. For several input cases, \textit{WSUM} yields MSE that are orders of magnitude larger than that of DFAOIT and therefore was not included in the corresponding figures. 

\textit{WAVG} offers a better OIT quality as it distributes the transmittance equally among the transparent fragments. It alleviates the color saturation exhibited by \textit{WSUM} and uses the exact transmittance product for the background, but comes with poor transparency results, as it fails to take into account object occlusion.

\textit{HT} increases quality by combining both exact and approximate OIT methods. It computes the exact OIT color of the $k$ closest fragments and approximates the OIT color (WAVG) for the remaining fragments. We used HT with $k=2$ as our testing scenario, as our method similarly captures the exact color of the first two fragments. This aims to show that our method does not just blend the exact and approximate parts, but inferes a better OIT approximation.

\textit{WBOIT} further improves visual quality by extending the blending method by using weights, that decrease with the distance to the camera, for the colors. Weights create an occlusion cue between transparent surfaces and work as an occlusion estimator. However, \textit{WBOIT} weights are scene and camera specific and may require fine-tuning, a cumbersome process that has a significant impact on the final color, and does not always provide optimal results. 

\textit{MBOIT} achieves better results by replacing the empirically determined weights of WBOIT with an approximated transmittance, that is captured and reconstructed using moments. We use four power moments as our test scenario, however several variants (number of power moments, trigonometric moments) were also introduced by their authors, each with trade-offs between quality, performance, and memory. The four moments version used here provides comparable time performance and memory requirements as our DFAOIT method.

Finally, our \textit{DFAOIT} outperforms the aforementioned approaches due to its ability to determine a more sophisticated color composition, as the color blending is performed by the neural network. The neural network input features cover the entire pixel space and provide color and opacity information. The exact color of the two closest fragments improves quality and enables our method to retain better details (Figures \ref{fig:teaser} and \ref{fig:graph_qcomp1}), while the remaining input features improve the color approximation of the remaining $n-2$ fragments. Our method outperforms the competing methods in increasingly higher opacity scenes, as the closest fragment contributions become increasingly more important (Figures \ref{fig:graph_rel} and \ref{fig:graph_qcomp1}).

Our experimental evaluation demonstrates that our method outperforms the competing methods in a plethora of testing scenarios with a variety of scenes with several opacity and depth complexity values in terms of MSE as shown in Figures \ref{fig:graph} and \ref{fig:graph_rel}, including more complex and high depth complexity scenes (Fig \ref{fig:graph}, Hairball D: 120).

\begin{figure*}[h]
	\centering
	\includegraphics[width=0.65\textwidth]{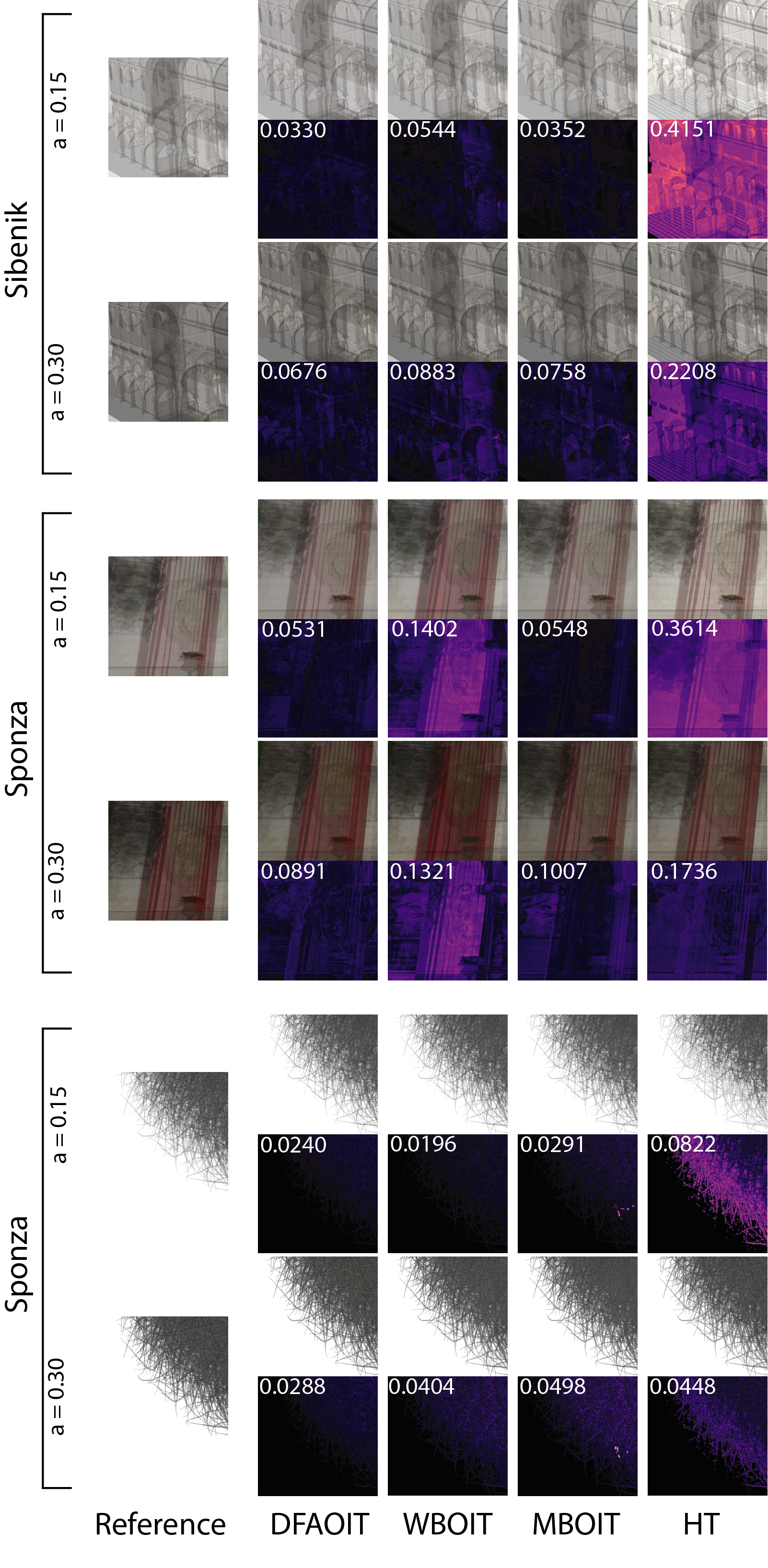}
	\caption{A quality comparison between the four approaches. An error map (generated using the \FLIP tool \cite{Andersson2020}) is computed between each method and the ground truth. On the left corner of each image is the \FLIP mean error, that certifies that our method displays better results compared to the other three approaches.}
	\label{fig:graph_scenes_qcomp1}
\end{figure*}

\begin{figure*}[h]
	\centering
	\includegraphics[width=\textwidth]{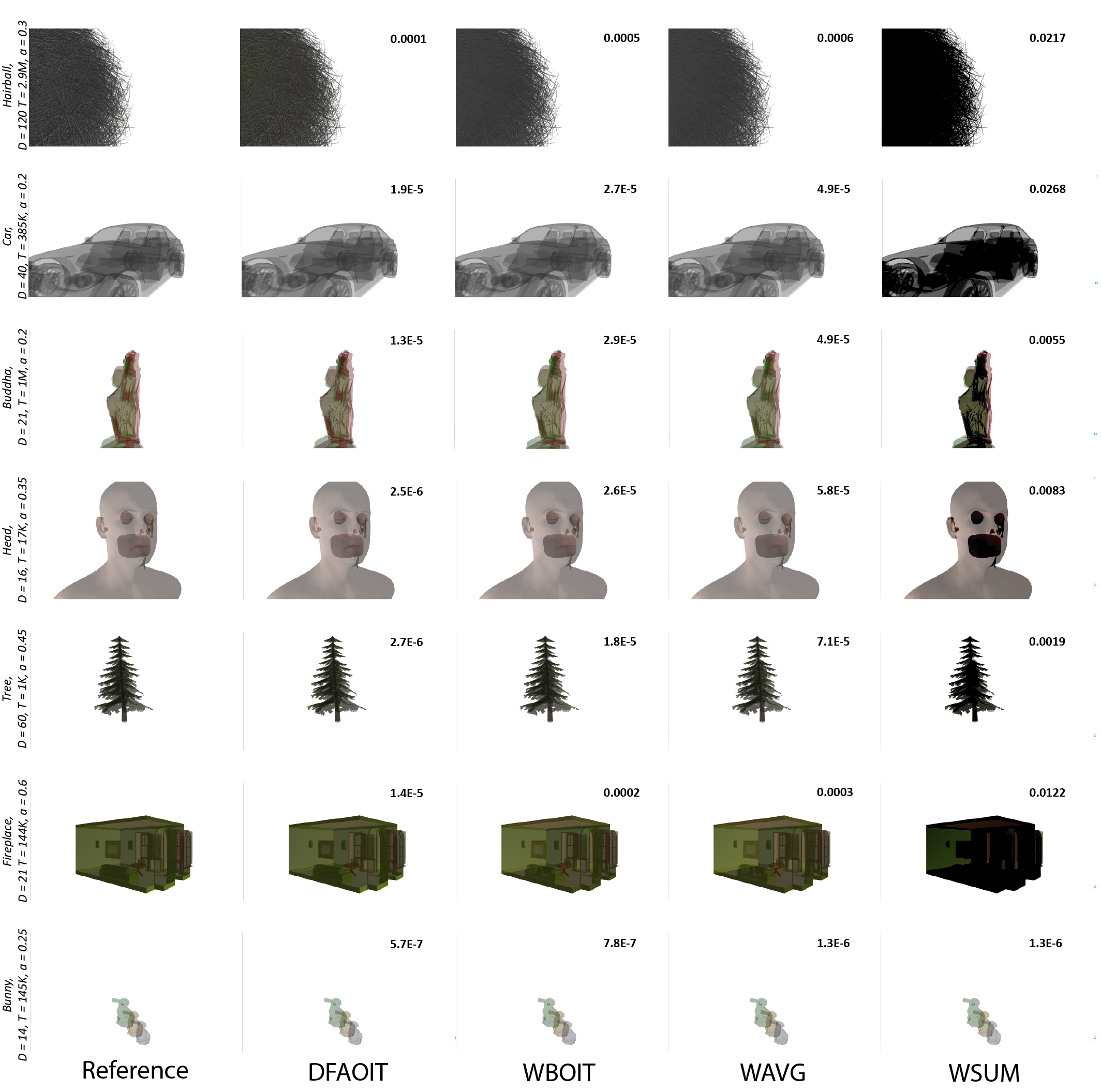}
	\caption{A quality comparison between the four approaches. On the top right of each image is the mean squared error.}
	\label{fig:graph_scenes_qcomp2}
\end{figure*}

Additionally, Figures \ref{fig:graph_scenes_qcomp1} and \ref{fig:graph_scenes_qcomp2} provide qualitative and quantitative results that prove that our method provides better accuracy in terms of both the \FLIP measure and the MSE metric, as well as substantiating that our method offers a better approximation of the color of the remaining fragments compared to HT.

Finally, we have provided a video demonstration as supplementary material  
that shows real-time behavior of our method, while avoiding view-dependent artifacts during camera movement. The video also substantiates that we have developed a very efficient method that facilitates rendering of high depth-complexity scenes with quality very similar to the ground truth A-buffer result.
\section{Conclusions}
\label{sec:conclusions}
Our method employs machine learning and can be used to approximately render OIT in real time with high accuracy. The shader source code provided can be adapted for rendering transparent scenes in games, computer aided design systems, visualization platforms and virtual reality applications.

Without loss of generality, the neural network can be trained with perceptual image measures such as \FLIP \cite{Andersson2020} or with other transparency techniques besides OIT. As a future research direction we will investigate a neural network in a per-fragment basis that will convey fragment information in the sense of the blended OIT techniques. This approach may reduce the error in the expense of a small overhead in computation time and memory. 

\section*{Acknowledgements}
3D test models, except the ``Engine'' scene, were downloaded from Morgan McGuire's Computer Graphics Archive \cite{McGuire2017Data}.

\section*{Funding}
We acknowledge support of this work by the project “Dioni: Computing Infrastructure for Big-Data Processing and Analysis.” (MIS No. 5047222) which is implemented under the Action “Reinforcement of the Research and Innovation Infrastructure”, funded by the Operational Programme "Competitiveness, Entrepreneurship and Innovation" (NSRF 2014-2020) and co-financed by Greece and the European Union (European Regional Development Fund).


\clearpage
\bibliographystyle{ACM-Reference-Format}
\bibliography{dfa-oit}



\end{document}